# Shubnikov-de Haas quantum oscillations with large spin splitting in high-mobility $Al_{0.8}Ga_{0.2}Sb$/InAs/ $Al_{0.8}Ga_{0.2}Sb$ quantum-well heterostructures

Zhenghang Zhi,[1,b] Hanzhi Ruan,[1,b] Jiuming Liu,[1] Xinpeng Li,[1] Yong Zhang,[1] Qi Yao,[2,3] Chenjia Tang,[2,3] Yujie Xiao,[3] and Xufeng Kou [1,2,a]

[1] *School of Information Science and Technology, ShanghaiTech University, Shanghai, 201210, China*

[2] *ShanghaiTech Laboratory for Topological Physics, ShanghaiTech University, Shanghai 200031, China*

[3] *School of Physical Science and Technology, ShanghaiTech University, Shanghai, 201210, China*

______________________________

[a] Email: kouxf@shanghaitech.edu.cn

[b] Zhenghang Zhi and Hanzhi Ruan contributed equally to this work.

**Abstract:**

We report the epitaxial growth of high-quality $Al_{0.8}Ga_{0.2}Sb$/InAs/ $Al_{0.8}Ga_{0.2}Sb$ quantum well films featured by high carrier mobility and strong spin-orbit coupling. By appropriately optimizing the Al-to-Ga ratio in the AlGaSb barrier layer, the quantum confinement of the heterostructure is significantly enhanced, which results in both an ultra-high electron mobility of $9.24 \times 10^5 cm^2 \cdot V^{-1} \cdot s^{-1}$ and a giant magnetoresistance ratio of $3.65 \times 10^5$% at low temperatures. Meanwhile, pronounced Shubnikov-de Haas quantum oscillations persist up to 30 K, and their single-frequency feature indicates a well-defined Fermi surface without subband mixing in the two-dimensional electron gas channel. Moreover, the large effective g-factor of 12.93 leads to the observation of Zeeman splitting at large magnetic fields. Our results validate the AlGaSb/InAs quantum well heterostructures as a suitable candidate for constructing energy-efficient topological spintronic devices.

Benefiting from high electron mobility, strong spin-orbit coupling, small effective electron mass, and narrow direct bandgap, InAs and InSb are highly promising for high-speed electronic, spintronic, and infrared optoelectronic applications[1-3]. In particular, InAs/GaSb-based heterostructures have emerged as an intriguing platform, as their type-II band alignment gives rise to non-trivial topological edge states under an appropriate band-inversion configuration[4]. For instance, recent advancements in the fabrication of nano-scale InAs/GaSb quantum wells (QWs) have demonstrated the realization of quantum spin hall effect[5], manifesting the robust edge states protected by time-reversal symmetry(TRS)[3, 6]. On the other hand, the breaking of TRS can lead to an unconventional excitonic topological order in the heterostructure system under imbalanced electron and hole densities[7]. In addition to such unique topological features, the InAs/GaSb hetero-interfaces not only host a well-defined two-dimensional electron gas (2DEG) channel with an ultra-high carrier mobility, but also warrant the effective manipulation of spin states through the interfacial Rashba effect[8].

Despite the aforementioned attributes, the practical implementation of InAs/GaSb heterostructures faces several challenges, primarily due to the narrow bandgap nature of GaSb, which cannot provide an insulating barrier layer and invariably brings about leakage currents and reduced thermal stability. [9]. Alternatively, previous studies have explored several ternary or quadruple compound semiconductors (*e.g.* InGaAs, InAlAs, InGaAsSb, and InPAsSb) as a more viable buffer layer option in constructing high-performance InAs electronic device applications[10]. In this context, it is validated that the enlarged bandgap of AlGaSb ensures a superior quantum confinement of carriers within the InAs/AlGaSb interface, resulting in high mobility[11] and increased spin injection and detection efficiencies[12]. Moreover, by adjusting the Al and Ga composition, the lattice constant of AlGaSb can be finely tuned to achieve better lattice matching with InAs, thereby reducing interface defects[13].

In this work, we report the growth and characterization of high-quality $Al_{0.8}Ga_{0.2}Sb/InAs/Al_{0.8}Ga_{0.2}Sb$ quantum wells using molecular beam epitaxy (MBE). Guided by the band diagram of TCAD simulations, the optimized sample structure exhibits an ultra-high electron of $9.24 \times 10^5 cm^2 \cdot V^{-1} \cdot s^{-1}$ and Fermi velocity of $7.16 \times 10^5$ m/s, enabling the observations of giant magneto-resistance ratio above $10^5$ % and pronounced Shubnikov-de-Haas (SdH) quantum oscillations up to $T = 30$ K. Additionally, the angular dependence of the SdH oscillations reveals a large spin splitting with the g-factor of 12.93. Our findings demonstrate that optimizing the Al/Ga ratio in the AlGaSb barrier layer can effectively enhance the magneto-transport properties of the InAs/AlGaSb quantum-well system, and they also provide a foundation for the development of next-generation high-speed electronic and spintronic devices.

The goal of this work is to enhance the electrical performance of InAs/AlGaSb through quantum-well structural engineering, which includes the dedicated designs of the thickness and component ratio of the AlGaSb barrier layer. Accordingly, we first applied TCAD simulation (*i.e.*, which takes into account the polarization models, field-dependent mobility models and the Shockley-Read-Hall models) to investigate the band diagram of the InAs/AlGaSb QW heterostructures[14]. As illustrated in Fig. 1a, the optimized structure consists of an $Al_{0.8}Ga_{0.2}Sb(10$ nm$)/InAs(20$ nm$)/Al_{0.8}Ga_{0.2}Sb(20$ nm$)$ tri-layer configuration, with an Al-to-Ga ratio of 4 to ensure a well-defined QW while the lattice mismatch between InAs and AlGaSb is compromised. Based on the simulation results, high-quality $Al_{0.8}Ga_{0.2}Sb/InAs/Al_{0.8}Ga_{0.2}Sb$ thin films were grown on the *n*-type conductive GaSb(100) substrate by MBE. In particular, the GaSb substrate was pre-annealed at 530°C under a Sb-rich environment to remove the native oxide layer, and the pre-anneal process was completed by observing the $(2 \times 6)$ surface re-construction by an *in-situ* reflective high-energy electron diffractometer (RHEED) pattern. Afterwards, a 1 μm-thick insulating AlGaAsSb (*i.e.*, $E_g$=1.5eV) was introduced as the buffer layer to electrically isolate the QW from the conductive substrate[15]. After buffer layer growth, the $Al_{0.8}Ga_{0.2}Sb(10 \text{ nm})/InAs(20 \text{ nm})/Al_{0.8}Ga_{0.2}Sb$ (20 nm) QW structure was deposited in reference to the TCAD simulation. Finally, a heavily-doped InAs capping layer was grown to protect the AlGaSb barrier from oxidation as well as to facilitate Ohmic contact formation for device fabrication. It is worth noting that a 2D streaky RHEED pattern is found to persist during the entire growth, validating the epitaxial growth mode without involving the lattice mismatch-induced macro-structural defects. Concurrently, the cross-sectional transmission electron microscopy (TEM) image also manifests the absence of macroscopic threading dislocations in the well-established QW structure with sharp interfaces. After sample growth, we fabricated the InAs/AlGaSb quantum well film into μm-sized six-terminal Hall bar devices by a standard nano-fabrication process. As displayed in the inset of Fig.1b, the etching process was precisely controlled to stop at the insulating AlGaAsSb buffer layer to avert shorting with the conductive GaSb substrate. Meanwhile, low-resistance Ohmic contacts were defined by depositing Ti(20 nm)/Au(100 nm) using the electron-beam evaporator (e-beam) and rapid thermal annealing (RTA) methods.

Subsequently, temperature-dependent magneto-transport measurements were conducted on the InAs/AlGaSb QW-based devices where an AC current ($|I| = 1 \text{ uA}, f = 67 \text{ Hz}$) was applied along the $x$-axis, while longitudinal ($V_{xx}$) and transverse ($V_{xy}$) voltages were simultaneously recorded using the SR830 lock-in amplifier. The temperature-dependent resistance curve exhibits a typical metallic $R$-$T$ behavior, namely the channel resistance decreases successively as the base temperature drops from $T = 298 \text{ K}$ to 50 K (Fig. 1b), indicating the presence of the 2DEG channel in the InAs/AlGaSb QW heterostructure at low temperature[16, 17]. Concurrently, Fig.1c summarizes the evolutions of the carrier mobility ($\mu$, red triangles) and the 2D density ($n_{2D}$, blue stars) with respect to temperature. It is seen that the carrier density decreases by an order of magnitude from 298 K to 50 K, mainly due to the thermally suppressed impurity ionization rate in the bulk InAs layer[18, 19]. As the Hall-bar device further cools down below 50 K, the bulk conduction is almost frozen out, leaving the confinement of itinerant electrons within the 2DEG channel at both InAs/AlGaSb hetero-interfaces[20]. As a result, such 2DEG-dominated transport characteristics endow a dramatic enhancement of carrier mobility in the low temperature region. As long as the phonon scatterings (*e.g.*, polar acoustic/optical phonon scatterings) are suppressed, the channel mobility is found to increase from $6.77 \times 10^3$ cm$^2 \cdot$V$^{-1} \cdot$s$^{-1}$ (298 K) to $8.46 \times 10^5$ cm$^2 \cdot$V$^{-1} \cdot$s$^{-1}$ (50 K), which corresponds to a 125× enhancement. Notably, the mobility of our sample reached a peak value of $\mu = 9.24 \times 10^5$ cm$^2 \cdot$V$^{-1} \cdot$s$^{-1}$ at $T = 25$ K, and the saturation behavior at $T < 25$ K reflects that neutral impurity scattering dominates in the deep cryogenic temperature region. Quantitatively, the overall $\mu$–$T$ curve can be described by applying Matthiessen's rule[21]:

$$\mu = (1/\mu_{\text{NE}} + 1/\mu_{\text{POP}} + 1/\mu_{\text{PE}})^{-1}$$

where $\mu_{\text{POP}}$, $\mu_{\text{PE}}$ and $\mu_{\text{NE}}$ correspond to the contributions from the thermally-activated polar optical phonon scattering, polar acoustic phonon scattering, and neutral impurity scattering, respectively. Among them, polar optical phonon scattering and polar acoustic phonon scattering (*i.e.*, $\mu_{POP} \propto T^{-4.03}$ and $\mu_{PE} \propto T^{-1.5}$)are temperature-dependent[22, 23], whereas $\mu_{\text{NE}}$ is regarded as a temperature-independent term since it mainly rises from the static impurity scattering in the 2DEG channels. As highlighted in Fig. 1c, the fitted data (solid line) are closely aligned with the experimental data (triangle symbol), unveiling the physical origin of the high-mobility InAs/AlGaSb QW system.

To further investigate the electrical properties of our MBE-grown samples, high-magnetic-field measurements were conducted at various temperatures. As shown in Fig. 2a, the longitudinal resistance $R_{xx}$ exhibits a typical parabolic behavior in the low-magnetic field region, and a giant magneto-resistance ratio of MR $= \frac{\rho_{xx}(B) - \rho_{xx}(0)}{\rho_{xx}(0)} \times 100\%$ =3.64 $\times 10^5$ % is obtained at $B = 14$ T. Strikingly, thanks to the high-mobility nature of the InAs/AlGaSb QW samples, pronounced SdH quantum oscillations are observed from $T = 1.5$ K to $T = 30$ K (Fig.2). Given the correlation of $\tau = \mu/m^*$, such a high carrier mobility also warrants a long scattering time ($\tau$) in the 2DEG channels, which ensures that electrons can coherently complete several cyclotron orbits before being scattered (*i.e.*, $\omega_c \tau \gg 1$, where $\omega_c$ is the resonance frequency)[24]. Likewise, according to $\omega_c = eB/m^*$, the small effective mass of InAs is able to satisfy the $\hbar\omega_C \gg k_B T$ condition in the low-temperature region, which in turn contributes to a marked Landau-level (LL) energy gap splitting[25]. As a result, the initiation of the SdH oscillation of our device is found to occur at a relatively low magnetic field of $B = 2.5$ T at 1.5 K. In alignment with Fig.2a, the Hall resistance $R_{xy}$ also displays well-resolved quantum oscillations, with its peak and valley positions consistent with the $R_{xx}$ data (*i.e.*, yet due to the high carrier density, quantum Hall effects (QHE) are not observed in the present device without gate tuning). Besides, it is noted that a unique double-peak feature is observed in the high magnetic-field region (inset of Fig. 2b), which will be addressed in the following section.

Based on these SdH oscillation results, quantitative analyses were further carried out to extract associated electrical parameters. Fig.3a displays the temperature-dependent SdH oscillation signals after removing their parabolic MR backgrounds. It is seen that the oscillation peak and valley positions remain almost unchanged with temperature, indicating a stable carrier density and a well-defined Fermi surface[26]. This temperature-invariant periodicity also suggests the well-developed Landau levels unaffected by thermal broadening[27]. In the meantime, the amplitudes of the SdH oscillation curves stay around 300 Ω in the entire [1.5 K, 30 K] region, again epitomizing the high carrier mobility, long quantum coherence time, and negligible impurity/interface scattering characteristics of our MBE-grown InAs/AlGaSb QW sample. Moreover, taking advantage of both the robust quantum confinement and the salient interface quality, the supplementary fast Fourier transform (FFT) data in Fig. 3b reveal a single oscillating frequency of $f_{SdH}$ = 31.1 T, which indicates the absence of neither parallel conduction channels nor sub-band mixing [28]. According to the Onsager formula of $F = (\Phi_0/2\pi^2)S_F$ (*i.e.*, where $\Phi_0 = \frac{h}{2e}$)[29], the Fermi surface can be obtained as $S_F = 2.969 \times 10^{-3} Å^{-2}$, and the corresponding Fermi momentum is $k_F = \sqrt{s_F/\pi} = 0.0307 Å^{-1}$. Subsequently, Fig. 3c illustrates a negative correlation between the normalized SdH oscillation amplitude $\Delta R(T)/\Delta R(1.5K)$ and temperature. It is recalled that electrons would be thermally excited across adjacent Landau levels, and the occupation of these levels invariably becomes less distinct due to thermal smearing, thereby diminishing the amplitude of the SdH oscillation. Quantitatively, the extent of this thermal damping is explained by the Lifshitz-Kosevich (L-K) theory, where the amplitude of $\Delta R_{xx}(T)$ is given by[30]:

$$\Delta R_{xx}(T) \propto \frac{\lambda(T)}{sinh\,\lambda(T)}e^{-\lambda(T)}$$

where $\lambda(T) = \frac{2\pi^2 k_B T m_{cyc}}{\hbar eB}$ is the thermal factor. By fitting the measured curve of Fig. 3c, the effective mass of electrons and the Fermi velocity are extracted to be $m_{cyc} = 0.0497 m_e$ and $v_F = \frac{\hbar k_F}{m_{cyc}} = 7.155 \times 10^5$ m·s$^{-1}$.

To understand the origin of the SdH oscillations and gain insight into the band dispersion near the Dirac point, we constructed the Landau level fan diagram by plotting $1/B$ as a function of the Landau index. It is observed from Fig. 3d that the $\Delta R_{xx}(T)$ peak-and-valley positions (*i.e.*, the half-integer and the integer LL index) are linear with $1/B$, and the Landau index $n$-axis intercept $n_x$ is nearly zero. Combined with the expression of $\phi_B = \pi(1 - 2\gamma)$ where $\gamma = \frac{1}{2} - n_x = \frac{1}{2}$ [31, 32], the Berry phase ($\phi_B$) of the 2DEG channel is found to be zero in our AlGaSb/InAs/AlGaSb sample (*i.e.*, which is different from the GaSb/InAs QW counterpart with a topologically-nontrivial edge state and a non-zero Berry phase of $\pi$ [33]). In this regard, it is proposed that on the one hand, the introduction of Al enlarges the bandgap of the AlGaSb barrier layer, which enhances the quantum confinement to boost mobility. On the other hand, as the Al content increases, the type-II QW-induced band inversion may no longer hold, and the band alignment of InAs/AlGaSb gradually transits from type-II to type-I, as confirmed by the TCAD simulation data (Fig. 1a). Therefore, our results justify the importance of the barrier layer composition in determining the magneto-transport behaviors of the InAs-based QW systems.

In addition to SdH oscillations, it is known that the spin imbalance and electron–electron interactions in the presence of large perpendicular magnetic fields can lead to an enhancement of the effective g-factor[34]. To address this point, angular-dependent magneto-transport measurements were conducted, and the resulting SdH oscillation curves are summarized in Fig.4a. It is discovered that when the tilt angle ($\theta$) between the magnetic field and the *z*-axis (*i.e.*, normal to the sample plane) increases from 45° to 60°, the SdH oscillation curve experiences a 180° phase shift, namely its peaks and valleys are reversed in the same $\Delta R_{xx}$-$B\cos(\theta)$ diagram, as indicated by the red and blue arrows. Under such circumstances, by applying the coincidence method $g^* = \frac{m_e}{m_{cyc}}\cos\theta_0$ (*i.e.*, where $g^*$ is the effective g-factor, and $\theta_0$ is the critical coincidence angle in reference to the inversion of the SdH oscillation amplitude [35]), the effective g-factor of our MBE-grown InAs/AlGaSb QW sample is found to be $g^* = 12.93$ , in a good agreement with the previous reports[36, 37]. Along with the strong intrinsic spin-orbit coupling of InAs, such a large g-factor leads to an evident spin-splitting of the SdH oscillation in the high-field region. As highlighted by the red arrows in Fig. 4b, the appearance of the double peaks of the SdH oscillations reside at $B = 10.5$ T and 12.2 T, respectively. Given the relation between the LL energy and Zeeman splitting energy: $E_n = \hbar\omega_c\left(n + \frac{1}{2}\right) + \frac{1}{2}g^*\mu_B B$, the Zeeman splitting will be visible only when the Zeeman energy term $g\mu_B B$ becomes comparable to the Landau energy $\hbar\omega_c$[38]. Accordingly, this high-field Zeeman splitting feature can be attributed to the small effective mass, large effective g-factor and strong spin-orbit coupling in our AlGaSb/InAs QW heterostructures, thereafter allowing for efficient spin injection, manipulation and detection in spin-based applications[39].

In conclusion, we have demonstrated the growth of high-quality $Al_{0.8}Ga_{0.2}Sb/InAs/Al_{0.8}Ga_{0.2}Sb$ quantum well thin film. Driven by the high mobility and large effective g-factor, the AlGaSb/InAs QW samples exhibit both a giant magnetoresistance and robust SdH oscillations up to 30 K; meanwhile, the strong spin–orbit coupling is manifested by the pronounced Zeeman splitting at higher magnetic fields. Moreover, by tuning the Al-to-Ga ratio of the AlGaSb barrier, we have integrated strong SOC and high mobility in a single platform, paving the way for next-generation high-speed and energy-efficient spintronic devices. Besides, our work also shows that structural engineering provides an effective way to tailor the band alignment, which may enable the optimization of electrical parameters and/or topological orders for various applications.

**Figure Caption:**

FIG. 1. (a) Left: Cross-sectional transmission electron microscopy image shows the well-ordered crystalline configuration. All the interfaces are distinguished without notable defects. Right: The band diagram of the $Al_{0.8}Ga_{0.2}Sb/InAs/Al_{0.8}Ga_{0.2}Sb$ quantum well by TCAD simulation. (b) Temperature-dependent resistance of the MBE-gorwn AlGaSb/InAs/AlGaSb sample. Inset: Illustration of the six-terminal hall bar device (c) Temperature-dependent mobility (pink triangles) and carrier concentration (blue pentagrams) results of the AlGaSb/InAs/AlGaSb sample.

FIG. 2. (a) Temperature-dependent magneto-resistance results of the InAs/AlGaSb QW sample with pronounced SdH oscillations from 1.5 K to 30 K. Inset: Parabolic MR curve observed in the low magnetic-field region. (b) Temperature-dependent Hall resistance of InAs/AlGaSb QW sample. The peaks and valleys of $R_{xy}$ are consistent with the $R_{xx}$ data. Inset: Zeeman splitting in the high magnetic-field region.

FIG. 3. (a) Temperature-dependent SdH oscillation signals after removing the parabolic backgrounds. (b) Fast Fourier transform of the SdH oscillations with a single frequency of $f$ = 31.1 T. (c) Temperature-dependent normalized SdH oscillation amplitude (blue circles) and the fitting results (red dashed line) using the L-K theory. (d) Landau level index plot where integers and half integers are represented by blue circles and red squares, respectively.

FIG. 4. (a) Angular-dependent SdH oscillations with the 180° phase shift when $\theta \geq 60°$. (b) Zeeman splitting of the SdH oscillation curve featured by the double peaks (red arrows) in the high magnetic-field region.

**Figures:**

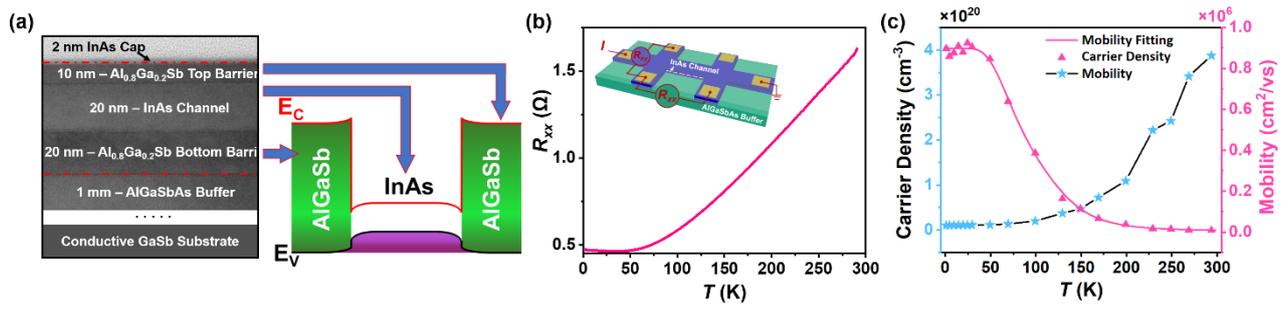

FIG. 1. Zhi *et al.*

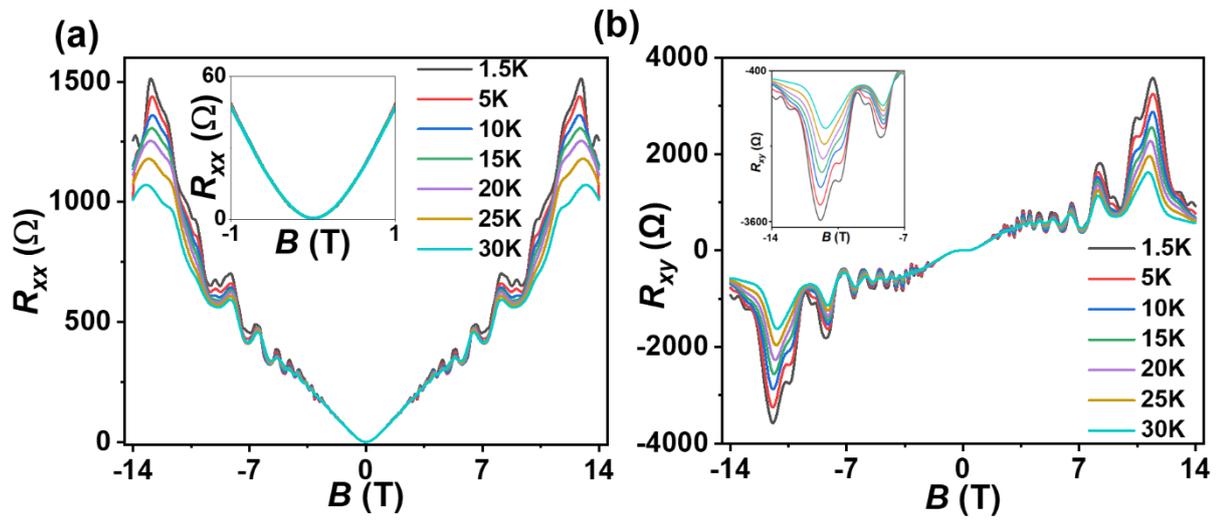

FIG. 2. Zhi *et al.*

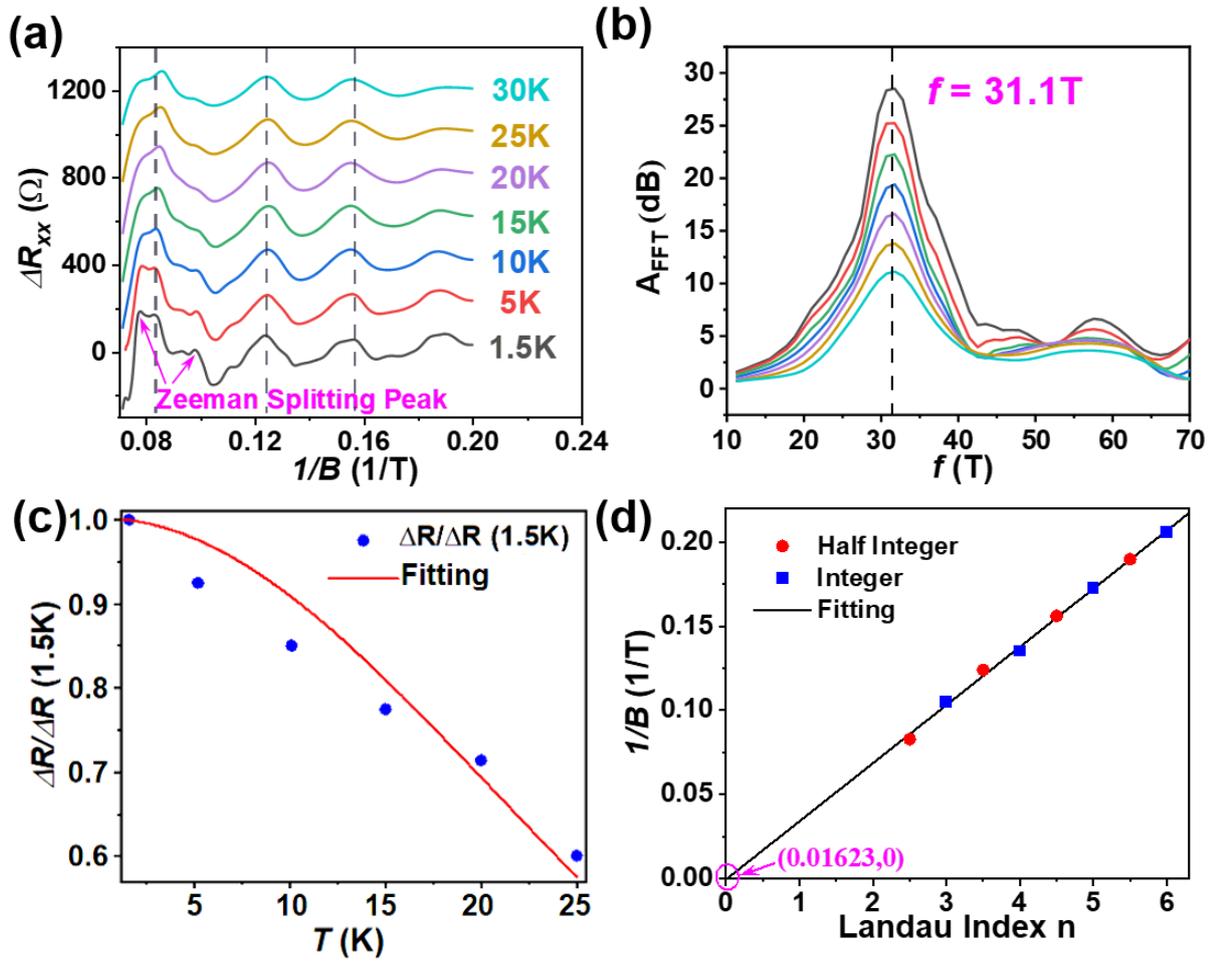

FIG. 3. Zhi *et al.*

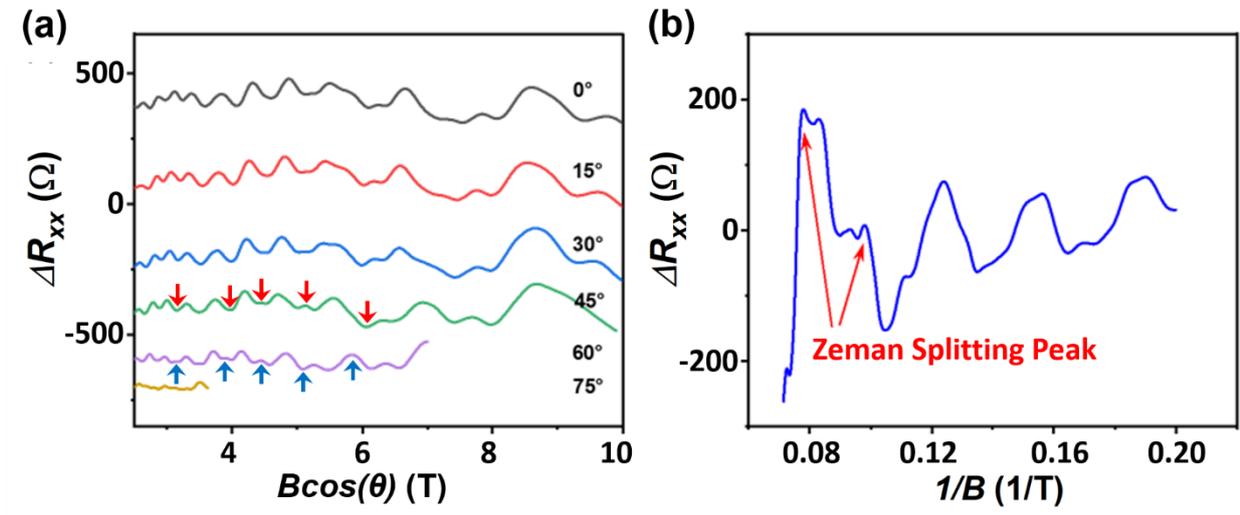

FIG. 4. Zhi *et al.*